\documentclass[12pt]{article}
\newcommand{\Comment}[1]{{}}
\usepackage[textwidth = 430 pt, textheight = 630 pt]{geometry}
\usepackage{amssymb,euscript,amsmath,amsfonts}

\usepackage{color}
\definecolor{MyDarkBlue}{rgb}{0.15,0.15,0.45}
 \usepackage[linktocpage=true]{hyperref}  
 \usepackage{graphicx}
\hypersetup{
colorlinks=true,
citecolor=MyDarkBlue,
linkcolor=MyDarkBlue,
urlcolor=MyDarkBlue,
pdfauthor={Neil Lambert },
pdftitle={ }, 
pdfsubject={hep-th}
}

\newcommand\ignore[1]{}
\def\one{{\,\hbox{1\kern-.8mm l}}}

\newcommand{\tr}{\operatorname{tr}}

\newcommand{\Cset}{{\,\,{{{^{_{\pmb{\mid}}}}\kern-.45em{\mathrm C}}}}}

\newcommand{\hG}{{\hat \Gamma}}

\newcommand{\nn}{\nonumber}

\newcommand{\be}{\begin{equation}}
\newcommand{\ee}{\end{equation}}
\newcommand{\bea}{\begin{eqnarray}}
\newcommand{\eea}{\end{eqnarray}}

\numberwithin{equation}{section}
\begin{document}

\renewcommand{\thefootnote}{\fnsymbol{footnote}}

\rightline{MTH-KCL/2017-03}

   \vspace{1.8truecm}

 \centerline{\LARGE \bf {\sc  The $(2,0)$ Superalgebra, Null M-branes}}
 \vskip 12pt
 
 \centerline{\LARGE \bf {\sc  and Hitchin's System} }
 
\vskip 2cm

  \centerline{
   {\large {\bf  {\sc P. Kucharski,${}^{\,a}$}}\footnote{E-mail address: \href{mailto:pjf.kucharski@fuw.edu.pl}{\tt pjf.kucharski@fuw.edu.pl}}     \,{\sc N.~Lambert$^{\,b}$}\footnote{E-mail address: \href{mailto:neil.lambert@kcl.ac.uk}{\tt neil.lambert@kcl.ac.uk}}  {\sc    and M.~Owen${}^{\,b}$}}\footnote{E-mail address: \href{mailto:miles.owen@kcl.ac.uk}{\tt miles.owen@kcl.ac.uk}}  }    
\vspace{1cm}
\centerline{${}^a${\it Faculty of Physics, University of Warsaw}}
\centerline{{\it ul. Pasteura 5, 02-093 Warsaw, Poland}} 
  
\vspace{1cm}
\centerline{${}^b${\it Department of Mathematics}}
\centerline{{\it King's College London, WC2R 2LS, UK}} 

\vspace{1.0truecm}

 
\thispagestyle{empty}

\centerline{\sc Abstract}
\vspace{0.4truecm}
\begin{center}
\begin{minipage}[c]{360pt}{
We present an interacting  system of equations with sixteen supersymmetries and an $SO(2)\times SO(6)$ R-symmetry where the fields depend on two space and one null dimensions that is derived from a representation of the six-dimensional $(2,0)$ superalgebra. The system can be viewed as two M5-branes compactified on $S^1_-\times {\mathbb T}^2$ or equivalently as M2-branes on ${\mathbb R}_+\times {\mathbb R}^2$,  where $\pm$ refer to null directions. We show that for a particular choice of fields the dynamics can be reduced to motion on the moduli space of solutions to the Hitchin system. We argue that this provides a description of intersecting null M2-branes and is also related by U-duality to a DLCQ description of four-dimensional maximally supersymmetric Yang-Mills.
 }

\end{minipage}
\end{center}

\newpage 
 
\renewcommand{\thefootnote}{\arabic{footnote}}
\setcounter{footnote}{0} 

\section{Introduction}

M-theory is generally viewed as a non-perturbative completion of string theory. While string theory is based on a perturbative quantization of strings there is no similar construction in M-theory. Branes of various types are known to play an important role in string theory. These often have a perturbative definition in terms of open strings and have decoupling limits leading to non-gravitating theories in any dimension less than ten. In M-theory one finds just M2-branes and M5-branes but embedded into an eleven-dimensional spacetime. Each of these admits a decoupling limit leading to interacting quantum field theories in three and six dimensions but there is typically no perturbative description (at least for smooth eleven-dimensional spacetime). These theories are of great interest as they are inherently strongly coupled and understanding them is thought to be a big step in the general understanding of M-theory. 

In  \cite{Lambert:2010wm,Lambert:2016xbs} a closed system of equations for various six-dimensional fields was obtained that are invariant under the $(2,0)$ superalgebra which is associated to the worldvolume of M5-branes embedded in an eleven-dimensional spacetime. The fields take values in a 3-algebra, except for the gauge field that takes values in the Lie-algebra (specifically $su(2)\oplus su(2)$ for the case at hand) that acts on the 3-algebra. The system can be thought of as a set of dynamical equations for the scalars, fermions and self-dual three-form as well as constraints for the additional gauge  and   vector fields that it contains. In addition the system depends on a choice of abelian three-form $C_{\mu\nu\lambda}$. For $C_{\mu\nu\lambda}=0$ it reproduces various descriptions of two M5-branes \cite{Lambert:2010wm,Lambert:2011gb,Hull:2014cxa}. For $C_{\mu\nu\lambda}$ spacelike the constraints reduce it to the equations two M2-branes \cite{Lambert:2016xbs}. 
The purpose of this paper is to explore the system for a null choice of  $C_{\mu\nu\lambda}$. We will see that this leads to a novel supersymmetric system of equations on ${\mathbb R}^2$ times a null direction ${\mathbb R}_+$. Alternatively, via an M-theory version of T-duality, we can think of this system as describing intersecting M2-branes which are tangent to a null direction.

A similar system of equations but defined on ${\mathbb R}^4$ times a null direction ${\mathbb R}_+$ was obtained in \cite{Lambert:2010wm} (and is therefore also a solution to the constraints of \cite{Lambert:2016xbs}). These were analysed in \cite{Lambert:2011gb} where it was shown they reduce to dynamics on instanton moduli space with the null direction playing the role of `time'. From the origin of these equations in the $(2,0)$ superalgebra it is clear that the resulting system describes two M5-branes compactified on a null circle with corresponding null momentum given by the instanton number. This is in agreement with the DLCQ prescription of \cite{Aharony:1997th,Aharony:1997an}. We similarly expect that the system here corresponds to two M5-branes compactified on ${\mathbb T}^2$ and carrying momentum along the null direction. We show that  the system reduces to quantum mechanics on Hitchin moduli space and provides a   description of intersecting null M2-branes. We note that there is a similar DLCQ description of four-dimensional maximally supersymmetric $SU(N)$ Yang-Mills with null momentum $K$  which is also based on quantum mechanics on Hitchin moduli space \cite{Ganor:1997jx,Kapustin:1998pb}. We will argue that this construction is related to our system by U-duality.

Another motivation for our work is to find and study field theories which have symmetry groups corresponding to branes embedded into eleven dimensions. From the field theory point of view   an embedding into eleven dimensions, as opposed to just ten, corresponds to enhanced R-symmetries, presumably arising at strong coupling. It is therefore of interest to obtain any such theories and study their interpretation: both as corresponding to objects in M-theory as well as strong coupling limits of field theories.

The rest of this paper is organised as follows. In section two we review the system of \cite{Lambert:2016xbs} and then examine it for the case of a null background 3-form $C_3$. In section three we analyse this new system and in particular show how, for a particular choice of fields,  it reduces to supersymmetric dynamics on the moduli space of solutions to Hitchin's equations. In section four we provide a physical interpretation of our system in terms of intersecting M2-branes. Section five contains our comments and conclusions on our results. We also provide an appendix with several conventions.

\section{The System}

Let us start by reviewing the $(2,0)$ system of \cite{Lambert:2016xbs} (which itself is a generalization of \cite{Lambert:2010wm}). The fields  $Y^\mu,X^i,H_{\mu\nu\lambda},\Psi$ all take values in a Lie-3-algebra, that is in a vector space  endowed with a totally anti-symmetric product $[\,\,\,,\,\,\,,\,\,\,]$ from the vector space to itself. Here $\mu,\nu=0,1,2,3,4,5$ and $i=6,7,8,9,10$.
If we expand all in fields in terms of a basis for the 3-algebra  $\{T^A\}$, {\it i.e.} $X=X_A T^A$, then  
\begin{equation}
[X,Y,Z]_D=X_AY_BZ_C f^{ABC}{}_{ D}\ ,
\end{equation}
where the structure constants of the 3-algebra $f^{ABC}{}_{ D}$ are anti-symmetric in the upper indices.  Furthermore the triple product is required to satisfy the  fundamental identity which reads
\begin{equation}
[U,V,[X,Y,Z]] = [[U,V,X],Y,Z]+ [X,[U,V,Y],Z]+ [X,Y,[U,V,Z]]\ ,
\end{equation}
or equivalently, the structure constants need to satisfy:
\begin{equation}
f^{[ABC}{}_{E} f^{D]EF}{}_{G}=0\ .
\end{equation}
We also require the existence of a symmetric inner-product which is invariant under the action of the 3-algebra, which allows the definition of a metric structure
\begin{equation}
h^{AB}=\langle T^A,T^B\rangle\ .
\end{equation}
This is equivalent to the condition $f^{[ABCD]}=0$, where $f^{ABCD}=f^{ABC}{}_Eh^{ED}$. In addition there is a gauge field $A_\mu$ which takes values in linear maps from the 3-algebra to itself and a covariant derivative:
\begin{equation}
D_\mu X_a=\partial_\mu X_a-(A_\mu)^b_{\,\,a} X_b=\partial_\mu X_a-A_\mu(X)_a\ .
\end{equation}
Lastly there is an abelian, constant, 3-form $C_{\mu\nu\lambda}$.

The  equations of motion are
 \begin{align}\label{eomfixed}
  0&=D^2 X^i-\frac{i}{2}[Y^\sigma,\bar \Psi,\Gamma_\sigma\Gamma^i\Psi]+[Y^\sigma,X^j,[Y_\sigma,X^j,X^i]]\nonumber\\&+\frac{i}{2\cdot3!}C^{\sigma\tau\omega}[\bar \Psi,\Gamma_{\sigma\tau\omega}\Gamma^{ij}\Psi,X^j]+\frac{1}{2\cdot 3! } C^{\sigma\tau\omega}C_{\sigma\tau\omega}[[X^i,X^j,X^k],X^j,X^k]\nonumber\\
 0 &= D_{[\lambda}H_{\mu\nu\rho]}+  \frac{1}{4}\varepsilon_{\mu\nu\lambda\rho\sigma\tau}[Y^\sigma,X^i,D^\tau X^i]- \frac{1}{2}(\star C)_{[\mu\nu\lambda}[X^i,X^j,[Y_{\rho]},X^i,X^j]]\nonumber\\
  &  + \frac{i}{8}\varepsilon_{\mu\nu\lambda\rho\sigma\tau}[Y^\sigma,\bar\Psi,\Gamma^{\tau}\Psi] - \frac{i}{2} (\star C)_{[\mu\nu\lambda}[X^i,\bar\Psi,\Gamma_{\rho]}\Gamma^i\Psi]\nonumber \\
 0&= \Gamma^\rho D_\rho\Psi + \Gamma_\rho\Gamma^i[Y^\rho,X^i,\Psi] + \frac{1}{2\cdot3!}\Gamma_{\rho\sigma\tau}C^{\rho\sigma\tau}\Gamma^{ij}[X^i,X^j,\Psi]\ ,
 \end{align}
 where $\Gamma^\mu,\Gamma^i$ are $32\times 32$ real $\Gamma$-matrices with $\mu,\nu,...=0,1,2,...,5$ and $i,j,...=6,7,8,9,10$. The spinors also satisfy
\begin{equation}
\Gamma_{012345}\epsilon = \epsilon\qquad \Gamma_{012345}\Psi = -\Psi\ ,
\end{equation}
and the three-form is self-dual:
\begin{equation}
H_{\mu\nu\lambda} = \frac{1}{3!}\varepsilon_{\mu\nu\lambda\rho\sigma\tau}H^{\rho\sigma\tau}\ .
\end{equation}
In addition to these equations of motion one has the constraints:
 \begin{align}\label{Constraintsfixed}
 0&= F_{\mu\nu}(\cdot) -[Y^\lambda,H_{\mu\nu\lambda},\ \cdot\ ] +(\star C)_{\mu\nu\lambda}[X^i,D^\lambda X^i, \ \cdot\ ] + \frac{i}{2}(\star C)_{\mu\nu\lambda}[\bar\Psi,\Gamma^\lambda\Psi, \ \cdot\ ]\nonumber\\
    0&=D_\nu Y^\mu-\frac{1}{2} \,C^{\mu\lambda\rho}H_{\nu\lambda\rho}\nonumber\\
    0&=C^{\mu\nu\sigma}D_\sigma (\cdot)+\,[Y^\mu,Y^\nu,\ \cdot\ ]\nonumber\\
      0&=[Y^\nu,D_\nu \cdot\ ,\ \cdot'\ ]+\frac{1}{3!}\,C^{\sigma\tau\omega}[H_{\sigma\tau\omega},\ \cdot\ ,\ \cdot'\ ]\nonumber\\
      0&=C\wedge Y\nn\\
      0 &= C_{[\mu\nu}{}^\rho C_{\lambda]\rho}{}^\sigma\ .
 \end{align}
 This system is invariant under the  supersymmetry transformations
  \begin{align} \label{Constraints}
  \delta X^i &= i\bar\epsilon \Gamma^i\Psi\nonumber\\
  \delta Y^\mu  &= \frac{i}{2} \bar\epsilon \Gamma_{\lambda\rho}C^{\mu\lambda\rho}\Psi \nonumber\\  
  \delta \Psi  &= \Gamma^\mu\Gamma^i D_\mu X^i\epsilon + \frac{1}{2\cdot 3!}H_{\mu\nu\lambda}\Gamma^{\mu\nu\lambda}\epsilon\nonumber\\&-\frac{1}{2}\Gamma_\mu\Gamma^{ij}[Y^\mu,X^i,X^j]\epsilon +\frac{1}{3!^2} C_{\mu\nu\lambda}\Gamma^{\mu\nu\lambda}\Gamma^{ijk}[X^i,X^j,X^k] \epsilon\nonumber\\
  \delta H_{\mu\nu\lambda}  &= 3i\bar\epsilon \Gamma_{[\mu\nu}D_{\lambda]}\Psi + i \bar\epsilon \Gamma^i\Gamma_{\mu\nu\lambda\rho}[Y^\rho,X^i,\Psi]\nonumber\\&  +\frac{i}{2} \bar\epsilon (\star C)_{\mu\nu\lambda}\Gamma^{ij}[X^i,X^j,\Psi] + \frac{3i}{4}\bar\epsilon \Gamma_{[\mu\nu|\rho\sigma}C^{\rho\sigma}{}_{\lambda]}\Gamma^{ij}[X^i,X^j,\Psi] \nonumber\\
 \delta A_\mu(\cdot)  &= i\bar\epsilon\Gamma_{\mu\nu}[Y^\nu,\Psi,\ \cdot\ ] -\frac{i  }{3!} \bar\epsilon C^{\nu\lambda\rho}\Gamma_{\mu\nu\lambda\rho}\Gamma^i[X^i,\Psi,\ \cdot\ ]\ .
 \end{align}

\subsection{A Null $C$ and $SO(2)\times SO(6)$}

In this paper we wish to analysis this system for the choice
 \begin{equation}\label{C}
C_{34+}=l^{3} ,
 \end{equation}
where 
\begin{align}
x^{+}&=\frac{x^{5}+ x^{0}}{\sqrt{2}}\qquad x^{-}=\frac{x^{5}- x^{0}}{\sqrt{2}}\ .
\end{align}

In particular we will see that the solution of the constraints leads to fields that only depend on $x^+,x^1,x^2$.  Although the system we started with has an $SO_L(1,5)\times SO_R(5)$ symmetry turning on $C_{+34}$ breaks the Lorentz group $SO_L(1,5)$ to an $SO_L(2)$ that acts as rotations in the $(x^1,x^2)$-plane along with an $SO_R(2)$ that acts as rotations in the $(x^3,x^4)$-plane and which is now viewed as an R-symmetry. Somewhat surprisingly we find that there is an enhancement of the original $SO_R(5)$ R-symmetry to $SO_R(6)$ so that the final system has an $SO_L(2)\times SO_R(2)\times SO_R(6)$ symmetry.

To exhibit this symmetry on the fermions it is useful to introduce a new representation of the $Spin(1,10)$ Clifford algebra:
\begin{align}
\hat{\Gamma}_0 &=  \Gamma_{0534}\nn\\
\hat{\Gamma}_{1,2} &= \Gamma_0 \Gamma_{1,2} \nonumber \\
\hat{\Gamma}_{3,4} &= \Gamma_{05} \Gamma_{4,3} \nonumber \\
\hat{\Gamma}^{5} &= \Gamma_0 \Gamma_{34} \nonumber \\
\hat{\Gamma}^i &= \Gamma_0 \Gamma^i \nonumber  \ ,
\end{align}
which satisfy $\{\hG_m,\hG_n\} = 2\eta_{mn}$, $m,n=0,1,2,..,10$.
However in what follows  we will only be interested in the $Spin(10)$ subalgebra which is broken to $Spin(2)\times Spin(2)\times Spin(6)$. We will also decompose any spinor $\chi$ as $\chi = \chi_++\chi_-$ where
\begin{equation}
\Gamma_{05}\chi_\pm  = \hG_{034}\chi=\pm\chi_{\pm} \ .
\end{equation}

\subsection{Solving the Constraints and Equations of Motion}

Our first task is to solve the constraints. From the last constraint in (\ref{Constraints}) we see that only $Y^-,Y^3,Y^4$ are non-vanishing. The third and fourth equations in (\ref{Constraints}) can be reduced to algebraic equations if we take
$\partial_-,\partial_3,\partial_4$ to vanish. Thus all fields are functions of $x^+,x^1,x^2$. Solving the resulting algebraic equations  from the third and fourth equations in (\ref{Constraints}) one finds that

\begin{align}
A_- &= \frac{1}{l^3} \left[ Y^3, Y^4, \cdot \right]\nonumber \\
A_3 &= \frac{1}{l^3} \left[ Y^4, Y^-, \cdot \right]\nonumber \\
A_4 &= -\frac{1}{l^3} \left[ Y^3, Y^-, \cdot \right] \ .
\end{align}

Next we can use the second equation in (\ref{Constraints}) to determine the components of $H_{\mu\nu\lambda}$. Using self-duality we find
\begin{align}
H_{34-}&=H_{12-}= -\frac{1}{l^{6}}\left[Y^{3},Y^{4},Y^{-}\right]\nn\\
H_{34+}&=-H_{12+}=\frac{1}{l^{3}}D_{+}Y^{-}\nn\\
H_{3-+}&=H_{124}=-\frac{1}{l^{3}}D_{+}Y^{4}\nn\\
H_{4-+}&=-H_{123}=\frac{1}{l^{3}}D_{+}Y^{3}\nn\\
H_{134}&=-H_{2-+}=\frac{1}{l^{3}}D_{1}Y^{-}\nn\\
H_{234}&=H_{1-+}=\frac{1}{l^{3}}D_{2}Y^{-}\nn\\
-\frac{1}{l^{3}}D_{1}Y^{4}=H_{13-}&=-H_{24-}=-\frac{1}{l^{3}}D_{2}Y^{3}\nn\\
\frac{1}{l^{3}}D_{1}Y^{3}=H_{14-}&=H_{23-}=-\frac{1}{l^{3}}D_{2}Y^{4}\ .
\end{align}

To proceed it is useful to introduce the complex coordinates and fields
\begin{align}
z & =x^{1}+ix^{2}\qquad \bar z =x^{1}-ix^{2}\nn\\
Z &=Y^{4}+iY^{3}\qquad \bar Z =Y^{4}-iY^{3} \ .
\end{align}
Here, and in what follows,  a bar denotes complex conjugation and not the Dirac conjugate. 
In addition we introduce an $SO(6)$ multiplet of scalar fields $X^I$, $I=5,6,...,10$, defined by
\begin{align}
X^5 = l^{-3}Y^- \qquad X^I = X^i\ I=6,...,10\ .
\end{align}

We first note that there is one independent  component of $H_{\mu\nu\lambda}$ that is  not determined from the constraints above and so we define
\begin{equation}
H = H_{+z3}=iH_{+z4}\ .
\end{equation}
We then find that the self-dual conditions $H_{13-}=-H_{24-}$ and $H_{14-}=H_{23-}$ are equivalent to 
\begin{equation}
\bar DZ=0\ .
\end{equation}
The remaining constraints can now be evaluated to give 
\begin{align}
 F_{+z} (\cdot) &= il^3  \left[ X^I, D X^I, \cdot \right] - i \left[ Z, H,\ \cdot \ \right] - \frac{l^3}{2} \left( \left[\Psi^T_+, \hat{\Gamma}_z \Psi_-, \cdot  \right] + \left[\Psi^T_-, \hat{\Gamma}_z \Psi_+, \cdot  \right]  \right) \nn\\
F_{z \bar{z}} (\cdot) &= -\frac{i}{4l^3} \left( \left[Z, D_+ \bar{Z},\ \cdot\  \right] + \left[ \bar{Z}, D_+ Z, \cdot \right] \right)   - \frac{1}{4 }   \left[ X^I, \left[Z, \bar{Z}, X^I  \right],  \ \cdot\ \right] - \frac{l^3}{2\sqrt{2}} \left[ \Psi_+^T, \Psi_+, \cdot  \right] \ . \end{align}
Our last job is to evaluate the equations of motion. The scalar  equation becomes
\begin{align}
0&= 2(D\bar D + \bar D D)X^I +  \frac{ i}{2l^3} [D_+Z,\bar Z,X^I] + \frac{ i}{2l^3} [Z,D_+\bar Z,X^I] + \frac{i}{ l^3} [Z,\bar Z,D_+X^I]\nonumber \\
&  +\frac12 \left[ Z, X^J, [ \bar{Z},X^J,X^I ] \right]+  \frac12 \left[ \bar{Z}, X^J, [ Z   ,X^J,X^I ] \right]\ - l^3 \sqrt{2} \left[ \Psi_+^T, \hat{\Gamma}_{Z \bar{Z}} \hat{\Gamma}^{IJ} \Psi_+, X^J \right] \nonumber \\ & +\frac{i}{2} \left( \left[ Z, \Psi^T_+ ,  \hat{\Gamma}_{Z} \hat{\Gamma}^I \Psi_- \right] - \left[ Z, \Psi^T_- ,  \hat{\Gamma}_{Z} \hat{\Gamma}^I \Psi_+ \right] + \left[ \bar{Z}, \Psi^T_+ ,  \hat{\Gamma}_{\bar{Z}} \hat{\Gamma}^I \Psi_- \right] - \left[ \bar{Z}, \Psi^T_- ,  \hat{\Gamma}_{\bar{Z}} \hat{\Gamma}^I \Psi_+ \right]   \right)   ,
\end{align}
where the $I=5$ component actually arises from the $(DH)_{z\bar z +-}$ equation. 
The only other new equation that arises from the $(DH)_{\mu\nu\lambda}$ equation comes from the  $(DH)_{z\bar z +3}$ and $(DH)_{z\bar z +4}$ terms and gives 
\begin{align}\label{F1}
0 &= D_+^2Z+ il^3[Z, X^I, D_+X^I] - \frac{l^6}{2}[X^I,X^J,[X^I,X^J,Z]] +4l^3 D \bar{H}\nonumber \\ & + \frac{l^3}{\sqrt{2}•} \left[ Z, \Psi_-^T , \Psi_- \right]  + i l^6 \left( \left[ \Psi_+^T, \hat{\Gamma}_{\bar{Z}} \hat{\Gamma}^I \Psi_-, X^I \right] - \left[ \Psi_-^T, \hat{\Gamma}_{\bar{Z}} \hat{\Gamma}^I \Psi_+, X^I \right]  \right) .
\end{align}
The fermion equations are
 \begin{align}
 0 & = D_+\Psi_+ + \sqrt{2}\hat{\Gamma}_z \bar{D} \Psi_- + \sqrt{2}\hat{\Gamma}_{\bar{z}} D \Psi_- + i l^3 \hat{\Gamma}_{Z \bar{Z}} \hat{\Gamma}^{IJ} \left[ X^I, X^J, \Psi_+ \right] \nonumber \\ & + \frac{1}{\sqrt{2}} \hat{\Gamma}^I \hat{\Gamma}_Z \left[ Z, X^I, \Psi_- \right] + \frac{1}{\sqrt{2}} \hat{\Gamma}^I \hat{\Gamma}_{\bar{Z}} \left[ \bar{Z}, X^I, \Psi_- \right]    .
 \end{align}
 and 
\begin{align}
0 &= \sqrt{2} \hat{\Gamma}_z \bar{D} \Psi_+ + \sqrt{2} \hat{\Gamma}_{\bar{z}} D \Psi_+ - \frac{i}{2l^3} \left[ Z, \bar{Z}, \Psi_- \right] \nonumber \\ & - \frac{1}{\sqrt{2}} \hat{\Gamma}^I \hat{\Gamma}_Z \left[ Z, X^I, \Psi_+ \right] - \frac{1}{\sqrt{2}} \hat{\Gamma}^I \hat{\Gamma}_{\bar{Z}} \left[ \bar{Z} , X^I, \Psi_+ \right]    .
\end{align}
Here we see that the equations of motion have a natural $SO_L(2)\times SO_R(2)\times SO_R(6)$ symmetry. In particular the field $Y^-$ has enhanced the original $SO_R(5)$ to $SO_R(6)$. 

\subsection{Supersymmetry}

The supersymmetry transformations can also be expressed as

\begin{align}
\delta X^I =& i \epsilon^T_+ \hat{\Gamma}^I \Psi_- + i \epsilon^T_- \hat{\Gamma}^I \Psi_+\nn\\
\delta Z =& 2 \sqrt{2} l^3 \epsilon^T_+ \hat{\Gamma}_{\bar{Z}} \Psi_{+}
\nn\\
\delta \bar{Z} =& - 2 \sqrt{2} l^3 \epsilon^T_+ \hat{\Gamma}_{Z} \Psi_{+}
\nn \\
\delta A_{z} =&   \sqrt{2} l^3 \epsilon_+^T \hat{\Gamma}^I \hat{\Gamma}_z [X^I, \Psi_+, \cdot]  +   i \epsilon^T_- \hat{\Gamma}_z \hat{\Gamma}_{\bar{Z}} \left[ \bar{Z}, \Psi_+, \cdot  \right] - i \epsilon^T_+ \hat{\Gamma}_z \hat{\Gamma}_{Z} \left[ Z, \Psi_-, \cdot  \right]
\nn\\
\delta A_{\bar{z}} =& -\sqrt{2} l^3 \epsilon_+^T \hat{\Gamma}^I \hat{\Gamma}_{\bar{z}} [X^I, \Psi_+, \cdot] +   i \epsilon^T_- \hat{\Gamma}_{\bar{z}} \hat{\Gamma}_{Z} \left[ Z, \Psi_+, \cdot  \right] - i \epsilon^T_+ \hat{\Gamma}_{\bar{z}} \hat{\Gamma}_{\bar{Z}} \left[ \bar{Z}, \Psi_-, \cdot  \right]
\nn \\
\delta A_+ =&  \sqrt{2} i \epsilon^T_- \hat{\Gamma}_Z \left[ Z, \Psi_-, \cdot \right] + \sqrt{2} i \epsilon^T_- \hat{\Gamma}_{\bar{Z}} \left[ \bar{Z}, \Psi_-, \cdot \right] \nonumber \\  & + 2l^3 \epsilon^T_- \hat{\Gamma}_{Z \bar{Z}} \hat{\Gamma}^I \left[X^I, \Psi_+, \cdot  \right] - 2l^3 \epsilon^T_+ \hat{\Gamma}_{Z \bar{Z}} \hat{\Gamma}^I \left[X^I, \Psi_-, \cdot  \right]
\end{align}

for the bosons, and as

\begin{align}
\delta \Psi_+  
=&   \frac{i}{\sqrt{2}l^3} \hat{\Gamma}^I \left[ Z, \bar{Z}, X^I  \right] \epsilon_- - \frac{i}{l^3} \left( \hat{\Gamma}_Z D_+ Z -  \hat{\Gamma}_{\bar{Z}} D_+ \bar{Z}  \right) \epsilon_+ \nonumber \\ & - \frac{1}{2} \left( \hat{\Gamma}_Z \hat{\Gamma}^{IJ} \left[ Z, X^I, X^J  \right] + \hat{\Gamma}_{\bar{Z}} \hat{\Gamma}^{IJ} \left[ \bar{Z}, X^I, X^J  \right]  \right)\epsilon_+ \nonumber \\ &\  + 2 \left( \hat{\Gamma}_{\bar{z}} \hat{\Gamma}^I D X^I + \hat{\Gamma}_{{z}} \hat{\Gamma}^I \bar{D} X^I  \right) \epsilon_+ \nonumber \\ & + \frac{\sqrt{2}i}{l^3} \left(  \hat{\Gamma}_{\bar{z}} \hat{\Gamma}_Z D Z - \hat{\Gamma}_z \hat{\Gamma}_{\bar{Z}}\bar D\bar{Z}\right) \epsilon_-\nn\\
\delta \Psi_-  =& -\sqrt{2} \hat{\Gamma}^I D_+ X^I \epsilon_+ - \frac{ \sqrt{2}i l^3}{3}\hat{\Gamma}_{Z \bar{Z}} \hat{\Gamma}^{IJK} \left[ X^I, X^J, X^K  \right] \epsilon_+   \nonumber \\ & + \frac{1}{2} \left( \hat{\Gamma}_Z \hat{\Gamma}^{IJ} \left[ Z, X^I, X^J  \right] + \hat{\Gamma}_{\bar{Z}} \hat{\Gamma}^{IJ} \left[ \bar{Z}, X^I, X^J  \right]  \right)\epsilon_- \nonumber \\  &- \frac{i}{l^3} \left( \hat{\Gamma}_Z D_+ Z - \hat{\Gamma}_{\bar{Z}} D_+ \bar{Z}  \right) \epsilon_- \nonumber \\ & + 2 \left( \hat{\Gamma}_{\bar{z}} \hat{\Gamma}^I D X^I + \hat{\Gamma}_{{z}} \hat{\Gamma}^I \bar{D} X^I  \right) \epsilon_- \nonumber \\ & +2 \sqrt{2}i \left( \hat{\Gamma}_{\bar{z}}\hat{\Gamma}_{\bar{Z}} H - \hat{\Gamma}_z  \hat{\Gamma}_Z \bar{H} \right)  \epsilon_+\ 
\end{align}

for the fermions.

The variation of $H=H_{+z3}$ requires special attention as self-duality implies
that $H = iH_{+z4}$. Evaluating these gives
\begin{align}
\delta H_{+z3} =&   \sqrt{2} \epsilon^T_- \left( \hat{\Gamma}_Z  - \hat{\Gamma}_{\bar{Z}} \right) D\Psi_- + \epsilon^T_+ \hat{\Gamma}_z \hat{\Gamma}_Z D_+ \Psi_- + \epsilon^T_- \hat{\Gamma}_z \hat{\Gamma}_{\bar{Z}} D_+ \Psi_+ \nonumber \\ & + \frac{i}{2}l^3 \epsilon^T_- \hat{\Gamma}_z \hat{\Gamma}_{\bar{Z}} \hat{\Gamma}^{IJ} \left[ X^I, X^J, \Psi_+  \right] + \frac{i}{2}l^3 \epsilon^T_+ \hat{\Gamma}_z \hat{\Gamma}_Z  \hat{\Gamma}^{IJ} \left[ X^I, X^J, \Psi_-  \right]\nn\\
&+\sqrt{2}\epsilon_-^T\hG_z\hG_{Z\bar Z}\hG^I[Z+\bar Z,X^I,\Psi_-]\nn\\
i\delta H_{+z4} =
&     \sqrt{2} \epsilon^T_- \left( \hat{\Gamma}_Z  + \hat{\Gamma}_{\bar{Z}} \right) D\Psi_- + \epsilon^T_+ \hat{\Gamma}_z \hat{\Gamma}_Z D_+ \Psi_- -\epsilon^T_- \hat{\Gamma}_z \hat{\Gamma}_{\bar{Z}} D_+ \Psi_+ \nonumber \\ & - \frac{i}{2}l^3 \epsilon^T_- \hat{\Gamma}_z \hat{\Gamma}_{\bar{Z}} \hat{\Gamma}^{IJ} \left[ X^I, X^J, \Psi_+  \right] + \frac{i}{2}l^3 \epsilon^T_+ \hat{\Gamma}_z \hat{\Gamma}_Z  \hat{\Gamma}^{IJ} \left[ X^I, X^J, \Psi_-  \right]\nn\\
&-\sqrt{2}\epsilon_-^T\hG_z\hG_{Z\bar Z}\hG^I[Z-\bar Z,X^I,\Psi_-]\ .
  \end{align}
Demanding that these are equal gives the condition
\begin{equation}
\epsilon^T_- \left(\sqrt{2} \hat{\Gamma}_{\bar{Z}}    D\Psi_- - \hat{\Gamma}_z \hat{\Gamma}_{\bar{Z}} D_+ \Psi_+   -\frac{i}{2}l^3  \hat{\Gamma}_z \hat{\Gamma}_{\bar{Z}} \hat{\Gamma}^{IJ} \left[ X^I, X^J, \Psi_+  \right]   +\sqrt{2}\hG_z\hG_{Z\bar Z}\hG^I[Z ,X^I,\Psi_-]  \right)=0
\end{equation}
As required this vanishes as a consequence of the fermion equation (\ref{F1}). As a result we find
\begin{align}
\delta H =
&   \sqrt{2} \epsilon^T_-  \hat{\Gamma}_Z    D\Psi_- + \epsilon^T_+ \hat{\Gamma}_z \hat{\Gamma}_Z D_+ \Psi_-  \nn\\
& + \frac{i}{2}l^3 \epsilon^T_+ \hat{\Gamma}_z \hat{\Gamma}_Z  \hat{\Gamma}^{IJ} \left[ X^I, X^J, \Psi_-  \right] + \sqrt{2}\epsilon_-^T\hG_z\hG_{Z\bar Z}\hG^I[\bar Z,X^I,\Psi_-]\ .
\end{align}
It is worth commenting that the identification $H_{+z3}=iH_{+z4}$ maps the   $SO_R(2)$ action as rotation by $\theta$ on $x^3,x^4$ to the $U(1)$ action 
$H\to e^{i\theta}H$.

We also note that a rescaling of $l$ can be absorbed by a rescaling of $x^+$ and $H$. Henceforth we simply take $l=1$.

\subsection{Energy-Momentum and Superalgebra}

The general form for the supercurrent and energy-momentum tensor were given in \cite{Lambert:2016xbs} as:\begin{align}
S^\mu=&-2\pi i\langle D_\nu X^i ,\Gamma^\nu\Gamma^i\Gamma^\mu \Psi\rangle+\frac{\pi i}{3!} \langle H_{\sigma\tau\omega},\Gamma^{\sigma\tau\omega}\Gamma^\mu\Psi\rangle-\pi i\langle[Y_\nu,X^i,X^j],\Gamma^\nu\Gamma^{ij}\Gamma^\mu\Psi\rangle\nonumber\\ &+\frac{\pi i}{3\cdot 3!}C_{\sigma\tau\omega}\langle[X^i,X^j,X^k],\Gamma^{ijk}\Gamma^{\sigma\tau\omega}\Gamma^\mu\Psi\rangle\ ,
 \end{align}
 and\footnote{This corrects a misprint in the fermion kinetic term contribution to $T_{\mu\nu}$ that appears in \cite{Lambert:2016xbs}.} \begin{align}
T_{\mu\nu}=&2\pi\langle D_\mu X^i,D_\nu X^i\rangle-\pi\eta_{\mu\nu} \langle D_\lambda X^i,D^\lambda X^i\rangle+\pi\langle[X^i,X^j,Y_\mu],[X^i,X^j,Y_\nu]\rangle\nonumber\\
-&\frac{\pi}{2}\eta_{\mu\nu}\langle[X^i,X^j,Y_\lambda],[X^i,X^j,Y^\lambda]\rangle+\frac{\pi}{2}\langle H_{\mu\lambda\rho},H_\nu^{\,\,\lambda\rho}\rangle\nonumber\\
-& i\pi\langle \bar \Psi,\Gamma_\mu D_\nu\Psi\rangle +i\pi\eta_{\mu\nu}\langle\bar \Psi,\Gamma^\lambda D_\lambda \Psi\rangle-i\pi\eta_{\mu\nu}\langle[\bar \Psi,Y^\lambda,X^i],\Gamma_\lambda\Gamma^i\Psi\rangle\nonumber\\
&+\frac{\pi}{3!}\langle[X^i,X^j,X^k],[X^i,X^j,X^k]\rangle( \,C_{\mu\tau\omega}C_\nu^{\,\,\,\,\tau\omega}-\frac{1}{3!}\eta_{\mu\nu}C^2)\nonumber\\
&+\frac{\pi}{3!}C_{\mu\lambda\rho}(\star C)_{\nu}{}^{\lambda\rho}\langle[X^i,X^j,X^k],[X^i,X^j,X^k]\rangle-\frac{i\pi }{ 3!}\eta_{\mu\nu}C^{\sigma\tau\omega}\langle[\bar{\Psi},\Gamma_{\sigma\tau\omega}\Gamma^{ij}\Psi,X^i ],X^{j}\rangle\ .
\end{align}
Setting the fermions to zero we find that in the case at hand
\begin{align}
T_{--}&= {2\pi}  \langle DZ,\bar D\bar Z \rangle -\frac{\pi}{2 }\langle[Z,\bar Z, X^I ],[  Z,\bar Z,X^I ] \rangle \nn\\
& =  {\pi} \partial \langle Z,\bar D\bar Z \rangle+  {\pi} \bar\partial \langle \bar Z, D Z \rangle\nn\\
T_{-+}& = - {4\pi}  \langle DX^I,\bar D\bar X^I \rangle-\frac{\pi}{2}\langle[Z, X^I,X^J],[ \bar Z,X^I,X^J] \rangle -  {\pi}   \langle D_+ Z, D_+\bar Z  \rangle \nn\\
& = -2\pi \partial \left( \langle X^I, \bar{D} X^I \rangle +   \langle \bar{Z}, \bar{H} \rangle \right) - 2\pi \bar{\partial} \left( \langle X^I, D X^I  \rangle +  \langle Z, H \rangle   \right) \nn \\
& \phantom{=} - \frac{\pi}{2 } \partial_+ \left( \langle Z, D_+ \bar{Z} \rangle + \langle \bar{Z}, D_+ Z  \rangle \right) \nn \\
T_{-z} &= - {\pi}  \partial \langle Z, D_+ \bar{Z} \rangle \ .
\end{align}

In the system here the role of time is played by $x^+$ so we define 
\begin{align}
{\cal P}_+ &= V_3\int dzd\bar z\  T_{-+}\nn\\
{\cal P}_z & = V_3\int dzd\bar z\   T_{-z}\nn\\
{\cal Q}_\pm & = V_3\int dzd\bar z \  S^+_{\pm}\ ,
\end{align}
as well as the topological term
\begin{align}
{\cal W} &=V_3\int dzd\bar z\  T_{--} \ .
\end{align}
Here $V_3$ is a three-dimensional volume factor that arises from the fact that $T_{\mu\nu}$, as defined above, has dimension six as appropriate for a six-dimensional theory.
Given that there is only one length scale in our system it is natural to take $V_3= l^3$.
After some calculations one finds that the superalgebra takes the form
\begin{align}
\{ {\cal Q}_-,{\cal Q}_-\}   = &  2\sqrt{2}{\cal W} \nn\\
\{{\cal Q}_+,{\cal Q}_-\}  = & -4  {\cal P}_z \hat {\Gamma}_{\bar{z}} -4  {\cal P}_{\bar{z}} \hat{\Gamma}_z\nn\\ 
&  +4 {\cal Z}_Z^I \hat{\Gamma}_{\bar{Z}} \hat{\Gamma}^I + 4   {\cal Z}^I_{\bar{Z}} \hat{\Gamma}_Z \hat{\Gamma}^I\nonumber \\
&+ \frac{1}{2!}{\cal  Z}^{IJ}_{\bar z}\hat{\Gamma}_z \hat{\Gamma}^{IJ } +\frac{1}{2!}  {\cal  Z}^{ IJ}_z\hat{\Gamma}_{\bar z} \hat{\Gamma}^{IJ } 
\nn\\
&+ \frac{1}{3!}{\cal Z}^{IJK}_{\bar Z}\hat{\Gamma}_Z \hat{\Gamma}^{IJK} + \frac{1}{3!}{\cal Z}^{ IJK}_Z\hat{\Gamma}_{\bar Z} \hat{\Gamma}^{IJK} \nn\\
\{{\cal Q}_+,{\cal Q}_+\}   =&  -2\sqrt{2}{\cal P}_+ \nn\\
&+\frac{1}{2!}{\cal Z}^{IJ}_{z\bar{z}} \hat{\Gamma}_{z\bar{z}}
\hat{\Gamma}^{IJ}+\frac{1}{2!}{\cal Z}_{ Z\bar{Z}}^{IJ}\hat{\Gamma}_{Z\bar{Z}}\hat{\Gamma}^{IJ}\nonumber \\
 & +\frac{1}{3!}{\cal Z}^{IJK}_{\bar{Z}z}\hat{\Gamma}_{Z \bar{z}}\hat{\Gamma}^{IJK}+\frac{1}{3!}{\cal Z}^{IJK}_{\bar{Z}\bar{z}  }\hat{\Gamma}_{Zz} \hat{\Gamma}^{IJK} \nonumber \\
 & +\frac{1}{3!}{\cal Z}^{IJK}_{  Z z} \hat{\Gamma}_{\bar{Z}\bar{z}}\hat{\Gamma}^{IJK}+\frac{1}{3!}{\cal Z}^{IJK}_{Z\bar{z}}\hat{\Gamma}_{\bar{Z}z }\hat{\Gamma}^{IJK}+\frac{1}{4!}{\cal Z}^{IJKL}\hat{\Gamma}^{IJKL}\ .
\end{align}
The central charges are given by
\begin{align}
{\cal Z}_Z^I =& \quad {2\pi} i V_3 \int dzd\bar z\partial \langle X^I, \bar{D} \bar{Z} \rangle \nn \\
{\cal Z}_{\bar{Z}}^I  =& -2\pi i V_3 \int dz d\bar{z} \bar\partial \langle X^I, D Z  \rangle \nn \\
{\cal Z}_{\bar{z}}^{IJ} =& \quad 4 \pi i V_3 \int dz d\bar{z} \left( \langle  \bar{D}\bar{Z}  ,\left[ Z, X^I, X^J \right]  \rangle - 2 \langle \bar{D} X^I, \left[ Z, \bar{Z}, X^J \right]  \rangle \right) \nn \\
{\cal Z}_z^{IJ} =& -4 \pi i V_3 \int dz d\bar{z} \left( \langle D Z  ,\left[ \bar{Z}, X^I, X^J \right]  \rangle + 2 \langle D X^I, \left[ Z, \bar{Z}, X^J \right]  \rangle \right) \nn \\
{\cal Z}_{\bar{Z}}^{IJK} =& \quad 6 i \pi V_3 \int dz d\bar{z} \langle \left[Z, X^I, X^J \right], \left[Z, \bar{Z}, X^K \right] \rangle \nn \\
{\cal Z}_{Z}^{IJK} =& \quad 6 i \pi V_3 \int dz d\bar{z} \langle \left[\bar{Z}, X^I, X^J \right], \left[Z, \bar{Z}, X^K \right] \rangle \nn \\ \nn  
{\cal Z}_{z\bar{z}}^{IJ} =& -32 \sqrt{2} \pi V_3 \int dz d \bar{z} \langle D X^I, \bar{D} X^J \rangle \nn \\
{\cal Z}_{Z \bar{Z}}^{IJ} =& \quad 4 \sqrt{2} \pi V_3  \int dz d\bar{z} \left( 2\langle \left[ Z, X^I, X^K \right], \left[ \bar{Z}, X^J, X^K \right]  \rangle\right.\nn \\ &
\qquad\qquad\qquad \qquad   \left.+ i \langle \left[ Z, X^I, X^J  \right], D_+ \bar{Z} \rangle + i \langle \left[\bar{Z}, X^I, X^J \right], D_+ Z \rangle    \right) \nn \\
{\cal Z}_{\bar{Z}z}^{IJK} =& \quad 24 \sqrt{2} \pi V_3 \int dz d\bar{z} \langle \left[ Z, X^I, X^J \right], DX^K \rangle \nn \\
{\cal Z}_{\bar{Z}\bar{z}}^{IJK} =& \quad 24 \sqrt{2} \pi V_3 \int dz d\bar{z} \langle \left[ Z, X^I, X^J \right], \bar{D} X^K \rangle \nn \\
{\cal Z}_{Zz}^{IJK} =& \quad 24 \sqrt{2} \pi V_3  \int dz d\bar{z} \langle \left[ \bar{Z}, X^I, X^J \right], D X^K \rangle \nn \\
{\cal Z}_{Z\bar{z}}^{IJK} =& \quad 24 \sqrt{2} \pi V_3 \int dz d\bar{z} \langle \left[ \bar{Z}, X^I, X^J \right], \bar{D} X^K \rangle \nn \\
{\cal Z}^{IJKL} =& - 12 \sqrt{2}\pi V_3 \int dz d \bar{z} \langle \left[ Z, X^I, X^J  \right], \left[ \bar{Z}, X^K, X^L \right] \rangle \ ,
\end{align}
where anti-symmetrization on all free $I,J,K,L$ indices is understood.

\section{Reduction to Dynamics on Moduli Space}

We now turn to an analysis of the dynamical equations that we found above. We view $x^+$ as `time' and take the Hamiltonian to be $-{\cal P}_+$. They are a novel system of differential equations for a set of three-algebra valued fields $(X^I,Z,H,\Psi_+,\Psi_-)$ along with a Lie-algebra valued gauge field $(A_+,A_z,A_{\bar z})$ all of which depend on  two space and one null directions $(z,\bar z , x^+)$ and are invariant under 16 supersymmetries generated by ${\cal Q}_+$ and ${\cal Q}_-$.
  
\subsection{Abelian Case}

To gain some insight it is helpful to first solve the abelian case where the triple product vanishes and we set the gauge fields to zero. The equations of motion  are simply
\begin{align}
\partial_+\Psi_+ +\sqrt{2}\hG_z\bar\partial \Psi_- +\sqrt{2}\hG_{\bar z}\partial \Psi_- &=0\nonumber\\
\sqrt{2}\hG_z\bar\partial \Psi_+ +\sqrt{2}\hG_{\bar z}\partial \Psi_+ &=0\nn\\
\bar \partial Z & = 0\nn\\
\bar\partial\partial X^I & = 0\nn\\
\partial^2_+{\bar Z} +  {4}\bar \partial H &=0\ .
\end{align}
The solutions to these equations are readily seen to be given by taking $Z$ to be an arbitrary $x^+$ dependent holomorphic function of $z$ and $X^I$ can be taken to be the real part of an   arbitrary $x^+$ dependent holomorphic function. For $H$ we find 
\begin{align}
  H &=h-\frac14 \int_0^{\bar z}  \partial_+^2 
{\bar Z}({\bar z'})d{\bar z'}\ .
\end{align}
where   $h$ is a holomorphic function   which also has an arbitrary dependence on  $x^+$. 
Looking at the fermions we find
\begin{align}
\Psi_+  & = \eta_++ {\bar \eta_+} \nn\\
\Psi_-& = \eta_- + {\bar\eta}_- -\frac{1}{\sqrt{2}}\int_0^{  z}\hG_{z} \partial_+ 
{ \eta}_+({ z'})d{ z'}-\frac{1}{\sqrt{2}}\int_0^{\bar z} \hG_{{\bar z}} \partial_+ 
{\bar\eta}_+({\bar z'})d{\bar z'} \ ,
\end{align}
where $\eta_\pm$ are spinors which satisfy
\begin{align}
\hat\Gamma_{\bar z}\eta_{\pm}=0\ .
\end{align}
and which are also holomorphic functions   and arbitrary functions of $x^+$.  

Thus the solution space is a set of holomorphic functions with arbitrary $x^+$-dependence. To recover some physics  
we note that for generic solutions the energy $P_+$ will diverge due to the poles in the holomorphic functions. Thus on physical grounds we should take all holomorphic functions to be constant. In this case $P_+$ will still diverge due to the integral over $z$ however we could imagine putting the theory on a torus, reducing the system to a quantum mechanical model. In that case global consistency  requires that
\begin{align}
\partial_+\Psi_+ =0\qquad \partial_+^2Z=0\ .
\end{align}
In this way we see the recover the familiar free-dynamics of $\Psi_+$ and $Z$, although the $x^+$ dependence of $X^I$, $H$ and $\Psi_-$ remain unconstrained. Looking that the on-shell  supersymmetry in this case we see that
\begin{align}
\delta \Psi_+ & = -i({\hG}_Z \partial_+ Z -  {\bar \hG_{\bar Z}}\partial_+ {\bar Z})\epsilon_+
\nn\\
\delta \Psi_- &=  \frac{i}{l^3} \left( \hat{\Gamma}_Z \partial_+ Z - \hat{\Gamma}_{\bar{Z}} \partial_+ \bar{Z}  \right) \epsilon_-    +2 \sqrt{2}i \left( \hat{\Gamma}_{\bar{z}}\hat{\Gamma}_{\bar{Z}} H - \hat{\Gamma}_z  \hat{\Gamma}_Z \bar{H} \right)  \epsilon_+\nn\\
\delta Z & = 2\sqrt{2}\epsilon_+^T{ \hG}_{\bar Z}\Psi_+\nn\\
\delta X^I & =  i \epsilon^T_+ \hat{\Gamma}^I \Psi_- + i \epsilon^T_- \hat{\Gamma}^I \Psi_+\nn\\
\delta H & =  \epsilon^T_+ \hat{\Gamma}_z \hat{\Gamma}_Z \partial_+ \Psi_-   \ .
\end{align}
Thus under ${\cal Q}_+$ $(Z,\Psi_+)$ and $(X^I,H,\Psi_-)$ form separate multiplets whereas under ${\cal Q}_-$ 
$(Z,\Psi_+)$ and $H$ are invariant but $(X^I,\Psi_-)$ transform into in $(\Psi_+,Z)$.

Even in the non-abelian case one sees that there are no standard kinetic terms for $X^I$, $H$ and $\Psi_-$. Indeed there are no $D_+$ derivatives on $H$ or $\Psi_-$ and $D_+$ only appears linearly on $X^I$ and within a triple product. Thus  we will interpret  $X^I$, $H$ and $\Psi_-$  as, possibly $x^+$-dependent, background fields. Given a particular choice of these fields as functions of $z$ and $x^+$ the equations of motion then determine the behaviour of $Z$ and $\Psi_+$.

\subsection{Vacua of the Non-Abelian System}

Next we look at the form of the supersymmetry algebra. Here one sees that ${\cal Q}_-$ is broken unless
\begin{align}
{\cal W}=0\ .
\end{align}
However this implies that $DZ=0$ and hence $F_{z\bar z}(Z)=0$. This effectively reduces the system back to the abelian case. Thus in what follows we assume that ${\cal Q}_-$ is broken and set $\epsilon_-=0$. We then wish to examine the system where only ${\cal Q}_+$ acts dynamically.    The role of ${\cal Q}_-$ can then be thought of as mapping between different backgrounds defined by choices of $X^I, H$ and $\Psi_-$. 

In this paper we will only consider backgrounds which preserve all of the ${\cal Q}_+$ supersymmetries.
In particular for a generic $\epsilon_+$ one sees that such   backgrounds are of the form $\Psi_-=0$, $H=0$ with $D_+X^I=0$ and $[X^I,X^J,X^K]=0$. Henceforth we will only consider such solutions. In this case the gauge fields are also invariant under ${\cal Q}_+$. Therefore the dynamical fields are $Z$ and $\Psi_+$. For simplicity we will also set $\Psi_+=0$ with the understanding that their dynamics can be recovered by applying the ${\cal Q}_+$ supersymmetry to the bosonic equations.

To begin we note that the ground states with ${\cal P}_+=0$ correspond to
\begin{align}
DX^I = 0\qquad [Z, X^I,X^J]=0\qquad D_+Z=0\ ,
\end{align}
 and such states are indeed invariant under ${\cal Q}_+$ and can have a non-vanishing ${\cal W}$. The equations of motion reduce to simply
 \begin{align}
 \bar D Z & =0 \nn\\
 F_{z \bar{z}} (\cdot) &=     - \frac{1}{4 }   \left[ X^I, \left[Z, \bar{Z}, X^I  \right], \cdot \right]\ .
 \end{align}
Since the $X^I$ are covariantly constant: $DX^I=\bar D X^I=0$ this equation is essentially just that of a Hitchin system \cite{Hitchin:1986vp} but in a three-algebra format as we now detail.

To continue we consider the specific case of a positive-definite 3-algebra with generators $T^A$, $A=1,2,3,4$ whose inner-product  is $\langle T^A,T^B\rangle = \delta^{AB}$ and triple product 
\begin{align}
[T^A,T^B,T^C] = \frac{2\pi}{k}\varepsilon^{ABCD}T^D\ ,
\end{align}
where $k$ is a constant (usually taken to be integer).
The gauge field takes values in  $so(4) = su(2)\oplus su(2)$ and the fields $X^I$ and $Z$ are in the vector of $SO(4)$. Solutions for $X^I$ that satisfy $[X^I,X^J,X^K]=0$ can be expanded in terms of two constant $SO(6)$ vectors $u^I,v^I$: 
\begin{align}
X^I = u^I T^3+v^I T^4\ .
\end{align}
For generic choices of $u^I$ and $v^I$ the gauge group is completely broken and the vacuum equations have no non-trivial solutions. In particular $Z$ is also restricted to lie  in the $T^3$ and $T^4$ directions of the 3-algebra  and the gauge field is locally flat. As with the abelian case above all the non-zero components of the fields are given by holomorphic functions. However demanding that ${\cal W}$ and ${\cal P}_+$ be finite requires that these holomorphic functions are constant and space is compactified.

However if we take all the $X^I$ to be aligned in the 3-algebra, say $X^I = v^IT^4$  then there is an unbroken $SO(3)$. If we expand  $Z=\sum Z_AT^A$  then $DX^I=\bar DX^I=0$ implies $\partial v^I=\bar\partial v^I=0$ and 
$A_{z4}{}^b=A_{za}{}^4=0$, $a,b,=1,2,3$. The  solutions are then given by 
\begin{align}\label{Hitchin}
\bar {\bf D}{\bf Z} &=0\nn\\
 {\bf F}_{z \bar{z}} &=      -\frac{\pi^2|v|^2}{k^2 }[{\bf Z},\bar {\bf Z}]  \ ,
\end{align}
where a  bold face indicates that the components are orthogonal to $T^4$ in the three-algebra and re-expressed as elements of the $SO(3)$ Lie algebra: $({\bf Z})^a{}_b = \varepsilon^{ca}{}_b Z_c$, $  {\bf D}{\bf Z} = \partial {\bf Z} - [ {\bf A},{\bf Z}]$. Furthermore $[\ ,\ ]$ is the usual Lie-bracket.\footnote{Note that our conventions for matrix multiplication are somewhat unusual: $(MN)^A{}_BX_A = M^C{}_BN^{A}{}_CX_A$} In other words bold-faced fields can be viewed as taking values in the unbroken $su(2)$ Lie algebra. This is precisely the Hitchin system for gauge algebra $su(2)$ \cite{Hitchin:1986vp}.  The equations of motion allow for $Z_4$ to be any holomorphic function but demanding that ${\cal W}$ is  finite implies that $Z_4$ is constant. Thus the vacuum solutions are in a one-to-one correspondence with solutions to the Hitchin system for $su(2)$. 

It is useful to recall here that the Hitchin system itself is the dimensional reduction of the the four-dimensional self-duality equations to two-dimensions. In particular let us define
\begin{equation}\label{4DA}
 {\bf A}_3 = \frac{2\pi |v|}{k}\frac{{\bf Z}-\bar {\bf Z}}{2i}\qquad{\bf A}_4 = \frac{2\pi |v|}{k}\frac{{\bf Z} +\bar {\bf Z} }{2}\ .
\end{equation}
Equation (\ref{Hitchin}) can then be written as (recall that $z = x^1+ix^2$)
\begin{align}
{\bf F}_{13} &=- {\bf F}_{24}\nn\\
{\bf F}_{23} &= {\bf F}_{14}\nn\\
{\bf F}_{12} & = {\bf F}_{34}\ ,
\end{align}
which are indeed the self-duality conditions and ${\cal W}$ is the dimensional reduction of instanton number and as such is no longer integer.

\subsection{Dynamical Evolution}

Next we allow for $x^+$ dependence and allow $Z$ to be dynamical, although we continue to restrict to the ${\cal Q}_+$ invariant sector: $D_+X^I=H =[X^I,X^J,X^K]=\Psi_-=0$. For simplicity we also set $\Psi_+=0$ with the understanding its dynamics can be restored using the ${\cal Q}_+$ supersymmetry. Keeping $X^I = v^IT^4$ and $Z_4=w$ this requires that $\partial_+v^I=0$ and  $A_{+}^a{}_4=-A_{+}^4{}_a=0$.  It is helpful then to rewrite the equations for the various remaining fields which we now express in their $su(2)$-valued form.

We start with the observation that $({\bf B})^b{}_c = \varepsilon^{ab}{}_cA^a_z{}_4$ is not necessarily zero since $DX^I$ need not vanish. This implies that the holomorphic constraint ${\bar D}Z=0$ leads to the equations
\begin{align}
\bar\partial w + \frac12{\rm tr}(\bar {\bf B}{\bf Z}) & =0\nn\\
\bar {\bf D}{\bf  Z} +\bar {\bf B} w & =0\ ,
\end{align}
for the $A=4$ and $A=a$ components respectively. Thus a non-zero $w$ and ${\bf B}$ lead to change in the holomorphic constraint on ${\bf Z}$. 

Next we recall that the Hitchin equation (\ref{Hitchin}) which arose from the $(C,D) = (c,d)$ component of the $F_{z\bar z}$ equation now becomes
\begin{align}\label{Hfull}
{\bf F}_{z\bar z} =  -\frac{\pi^2|v|^2}{k^2 }[{\bf Z},\bar {\bf Z}] + [{\bf B},\bar {\bf B}] + \frac{i}{4}\left(\frac{2\pi}{k}\right)(w{\bf D}_+\bar {\bf Z}+\bar w{\bf D}_+ {\bf Z} - \bar{\bf Z}\partial_+w  - {\bf Z}\partial_+\bar w)\ ,
\end{align} 
where
\begin{equation}
{\bf F}_{z\bar z}  =\partial \bar {\bf A} - \bar \partial {\bf A} - [{\bf A},\bar {\bf A}]\ .
\end{equation}
If we examine the $(C,D) = (c,4)$ component of the $F_{z\bar z}$ equation we find
\begin{equation}\label{Fc4}
{\bf D}\bar {\bf B} - \bar{\bf D}{\bf B}  = -\frac{i}{4 }\left(\frac{2\pi}{k}\right)\left([{\bf Z},{\bf D}_+\bar {\bf Z}]+[\bar {\bf Z},{\bf D}_+ {\bf Z}]\right)\ .
\end{equation}
From  the $F_{+z}$ equation we learn that
\begin{align}\label{F+z}
{\bf D}_+{\bf B} &=0\nn\\
\partial_+ {\bf A} - {\bf D}{\bf A}_+ & = -\frac{2\pi i}{k} |v|^2 {\bf B}\ ,
\end{align}
due to the  $(C,D) = (c,d)$ and $(C,D) = (c,4)$ components respectively. 
From the $(D\bar D+\bar D D )X^I$ equation we find
\begin{align}\label{veq}
\partial\bar\partial v^I + \frac{1}{2} {\rm tr }(\bar{\bf B}{\bf B})v^I & = 0\nn\\
({\bf D}\bar {\bf  B}+ \bar {\bf D}{\bf  B})v^I + 2{\bf B}\bar\partial v^I + 2\bar {\bf B}\partial v^I& = \frac{i}{4 }\left(\frac{2\pi}{k}\right)\left([{\bf Z},{\bf D}_+\bar {\bf Z}]-[\bar {\bf Z},{\bf D}_+ {\bf Z}]\right)v^I\ ,
\end{align}
arising from to the  $A=4$ and $A  =  a$ components respectively. 
Lastly we also simply find
\begin{align}
{\bf D}_+^2 {\bf Z}=0\qquad \partial_+^2w=0\ .
\end{align}
We see that non-vanishing ${\bf B}$ and $w$ lead to a $z$-dependent $v^I$ and hence to a modification of Hitchin's system.

Our approach here is to treat $X^I$ and hence $v^I$ as a background field.  Elementary manipulations of the first equation in (\ref{veq}) show that
\begin{equation}
\oint v^Id v^I = \int  \frac{1}{2} {\rm tr }({\bf B}^\dag{\bf B})|v|^2 + \int  |\partial v^I|^2 + |\bar\partial v^I|^2\ge 0\ .
\end{equation}
Thus if we are interested in solutions for which $v^I$ approaches a non-zero constant value at infinity plus subleading terms then the left hand side vanishes. Therefore ${\bf B}=0$ and $v^I$ is constant.  Let us first consider the case when $w=0$. We then see that  Hitchin's equation is preserved for all time. Thus any dynamical motion can only take place on the moduli space of solutions to Hitchin's system. In addition the remaining dynamical equations are 
\begin{align}
\label{dynamics}
[{\bf Z},{\bf D}_+\bar {\bf Z}]=0,\qquad  \partial_+ {\bf A} = {\bf D}{\bf A}_+\ ,\qquad  {\bf D}_+^2{\bf Z}=0 \ .
\end{align}

To understand these equations we recall that $({\bf A},{\bf Z})$ are required to solve the Hitchin equations for all $x^+$. Thus motion can only take place on the moduli space solutions so that under $x^+\to x^++\epsilon$, 
\begin{equation}
\delta {\bf A} = \partial_+ {\bf A} \epsilon\qquad \delta {\bf Z} = \partial_+ {\bf Z}\epsilon \ ,
\end{equation}
where $\delta {\bf A}$ and $\delta {\bf Z}$ are fluctuations of the solution to Hitchin's equations: {\it i.e.} solutions to the linearised Hitchin equations. In particular these linearised equations are
\begin{align}\label{Hlin}
{\bf D}  \partial_+  \bar {\bf A} - \bar {\bf D}   \partial_+ {\bf A} &=- \frac{\pi^2}{k^2}|v|^2\left([  \partial_+ {\bf Z}, \bar {\bf Z}]+[{\bf Z},  \partial_+\bar {\bf Z}]\right)\nn\\
\bar{\bf D}  \partial_+{\bf Z} - [  \partial_+ \bar {\bf A},{\bf Z}]&=0
\ .
\end{align}
Using the second equation  in (\ref{dynamics}) we see that\begin{align}\label{Hlin2}
{\bf D}  \partial_+  \bar {\bf A} - \bar {\bf D}   \partial_+ {\bf A} &=  ({\bf D}\bar{\bf D}-\bar {\bf D}{\bf D}){\bf A_+}\nn\\
& =  -[{\bf F}_{z\bar z},{\bf A}_+]\nn\\
&= \frac{\pi^2}{k^2}|v|^2[[{\bf Z},\bar {\bf Z}],{\bf A}_+]\nn\\
& = -\frac{\pi^2}{k^2}|v|^2([[{\bf A}_+, {\bf Z}],\bar {\bf Z}]+[{\bf Z},[{\bf A}_+,\bar {\bf Z}] ])\nn\\
&=- \frac{\pi^2}{k^2}|v|^2\left([  \partial_+ {\bf Z}, \bar {\bf Z}]+[{\bf Z},  \partial_+\bar {\bf Z}]\right)\ ,
 \end{align}
 where in the last line we used the first equation in  (\ref{dynamics}). 
Thus  (\ref{dynamics}) imply the first equation in (\ref{Hlin}). Using  (\ref{dynamics}) the second equation in (\ref{Hlin}) becomes simply
\begin{align} \label{Hlin3}
\bar{\bf D}{\bf D}_+{\bf Z} & = 0\ .
\end{align}
Thus the dynamical equations (\ref{dynamics}) along with (\ref{Hlin3}) describe motion on the Hitchin moduli space.

To continue we note that we do not want to consider motion that arises from gauge transformations: $\delta {\bf A} = {\bf D}\omega$, $\delta {\bf Z} = [\omega,{\bf Z}]$. Therefore we impose that the fluctuations are orthogonal to gauge transformations\footnote{This is just the reduction of the standard  instanton moduli space gauge fixing condition ${\rm tr}\int A_1\delta A_1 + ... +A_4\delta A_4$ for the four-dimensional gauge field defined in (\ref{4DA}).}:
\begin{equation}
-\frac12{\rm tr}\int dzd\bar z\ \left[ 2 \bar {\bf D} \omega  \delta {\bf A}  +  2  {\bf D} \omega\delta \bar {\bf A}  + \frac{2\pi^2}{k^2}|v|^2\left(  [\omega,\bar {\bf Z}] \delta {\bf Z}  +   [\omega,{\bf Z}] \delta \bar {\bf Z}] \right)\right] = 0\ .
\end{equation}
Integrating by parts and demanding that $\omega$ is arbitrary gives the condition
\begin{equation}
{\bf D}  \delta  \bar {\bf A} + \bar {\bf D}  \delta {\bf A} = \frac{\pi^2}{k^2}|v|^2\left([{\bf Z}, \delta \bar {\bf Z}]+[\bar {\bf Z},\delta {\bf Z}]\right)
 \ .
\end{equation}
Identifying $\delta {\bf A} = \partial_+ {\bf A}\epsilon$, $\delta {\bf Z} = \partial_+ {\bf Z}\epsilon $ and combining with the first equation in (\ref{Hlin}) gives the gauge fixing condition:
\begin{align}\label{ttt}
\bar {\bf D}  \partial_+{\bf A} &= \frac{\pi^2}{k^2}|v|^2[ {\bf Z},\partial_+ \bar {\bf Z}]\ , 
\end{align}
 or equivalently using (\ref{dynamics})
 \begin{equation}\label{Hlin4}
 \bar {\bf D} {\bf D}{\bf A}_+  = \frac{\pi^2}{k^2}|v|^2[ {\bf Z},[{\bf A}_+,\bar {\bf Z}]]\ .
 \end{equation}

Thus for the background $X^I=v^IT^4$, $Z_4=0$ the whole dynamical system is reduced to  motion on the moduli space of solutions to Hitchin's equations with the dynamical equations (\ref{dynamics}), (\ref{Hlin3}) and gauge fixing condition (\ref{Hlin4}).  The Hamiltonian is given  by ${\cal H} = -{\cal P}_+$ which in turn is simply that of a $\sigma$-model on the moduli space:
\begin{align}
{\cal H} & = \pi \int dzd\bar z \langle D_+Z,D_+{\bar Z}\rangle\nn\\
 & = -\frac{\pi}{2}{\rm tr} \int dzd\bar z \left((\partial_+{\bf Z} -[{\bf A_+},{\bf Z}] )(\partial_+\bar {\bf Z} -[{\bf A}_+,\bar {\bf Z}])\right)\nn\\
 &=- \frac{\pi}{2}{\rm tr} \int dzd\bar z \left( \partial_+{\bf Z} \partial_+\bar {\bf Z} - {\bf A}_+[\bar {\bf Z},\partial_+{\bf Z}] -   {\bf A}_+[ {\bf Z},\partial_+\bar{\bf Z}] + {\bf A}_+[ {\bf Z} ,[ {\bf A}_+,\bar {\bf Z}]]\right)\nn\\
  &=- \frac{\pi}{2}{\rm tr} \int dzd\bar z \left( \partial_+{\bf Z} \partial_+\bar {\bf Z}  -\frac12  {\bf A}_+[\bar {\bf Z},[{\bf A}_+, {\bf Z}]] -  \frac12 {\bf A}_+[ {\bf Z},[{\bf A}_+,\bar {\bf Z}]] \right)\nn\\
 & =  -\frac{\pi}{2} \int dzd\bar z \left(\partial_+{\bf Z}\partial_+\bar{\bf Z} -  \frac{k^2}{2\pi^2 |v|^2}{\bf A_+}{\bf D}\bar {\bf D}{\bf A_+}-  \frac{k^2}{2\pi^2 |v|^2}{\bf A_+}\bar{\bf D} {\bf D}{\bf A_+}\right)\nn\\
& = -\frac{k^2}{2\pi |v|^2}{\rm tr}\int dzd\bar z \left(   \frac{  \pi^2 |v|^2}{k^2}\partial_+{\bf Z} \partial_+\bar  {\bf Z}+  \partial_+{\bf A} \partial_+\bar  {\bf A} \right)\nn\\
& = \frac{k^2}{ 2\pi |v|^2}g_{mn}\partial_+ \xi^m\partial_+ \xi^n\ ,
\end{align}
where we have used the relations $[{\bf Z},\partial_+\bar {\bf Z}]= [{\bf Z},[{\bf A}_+,\bar {\bf Z}]]$,  $\bar {\bf D}  \partial_+{\bf A}  = \frac{\pi^2}{k^2}|v|^2[ {\bf Z},\partial_+ \bar {\bf Z}]$ and $\partial_+{\bf A} = {\bf D}{\bf A}_+$. Furthermore  
$\xi^m$ are the moduli space coordinates and  
\begin{align}
g_{mn} & =  -\frac{1}{2}{\rm tr}\int dzd\bar z \left( \delta_m{\bf A}_1\delta_n  {\bf A}_1+\delta_m{\bf A}_2\delta_n  {\bf A}_2+ \delta_m{\bf A}_3\delta_n{\bf A}_3+\delta_m {\bf A}_4\delta_n {\bf A}_4 \right)\ ,
\end{align}
is the natural metric on the moduli space. As shown by Hitchin \cite{Hitchin:1986vp} this space is hyper-Kahler and therefore, by standard arguments, the dynamics can be extended to include fermions in such a way as to preserve the 8 supersymmetries generated by ${\cal Q}_+$.

Next we can consider the effect of a non-zero $w$ but we still keep $v^I$ constant and hence ${\bf B}=0$. We see that for static solutions with $\partial_+ ={\bf A}_+=0$ we still reduce to Hitchin's system however for ${\bf A}_+,\partial_+ \ne 0$ there is a modifcation. To see what happens we can differentiate (\ref{Hfull}) with respect to $\partial_+$    to find (recall that ${\bf D}^2_+{\bf Z}=\partial_+^2w=0$):
\begin{align}
\bar {\bf D}\partial_+{\bf A} - \bar{\bf D} \partial_+{\bf A} 
&=  -\frac{\pi^2|v|^2}{k^2 }\left([{\partial}_+ {\bf Z},\bar {\bf Z}]+[{\bf Z},{\partial}_+\bar {\bf Z}]\right)\nn\\
&   + \frac{i}{4}\left(\frac{2\pi}{k}\right)(w\partial_+{\bf D}_+\bar {\bf Z}+\bar w\partial_+{\bf D}_+ {\bf Z} -   \partial_+w[{\bf A}_+,\bar{\bf Z}]  -  \partial_+{\bar w}[{\bf A}_+, {\bf Z}] )\nn\\
& =   \frac{\pi^2}{k^2}|v|^2[[{\bf Z},\bar {\bf Z}],{\bf A}_+]\nn\\
&   - \frac{i}{4}\left(\frac{2\pi}{k}\right)[w{\bf D}_+\bar {\bf Z} +{\bar w}{\bf D}_+ {\bf Z} -   \partial_+ w\bar{\bf Z}   -  \partial_+{\bar w} {\bf Z},{\bf A}_+] \ .
\end{align}
This generalises the first equation in (\ref{Hlin}) and the rest of the analysis continues as before. One sees that the analysis in (\ref{Hlin2}) still goes through one still finds that  (\ref{dynamics}) imply the first equation in (\ref{Hlin}). However (\ref{Hlin4}) is now modified to 
\begin{align} 
 \bar {\bf D} {\bf D}{\bf A}_+  &= \frac{\pi^2}{k^2}|v|^2[ {\bf Z},[{\bf A}_+,\bar {\bf Z}]]\nn\\
 &-  \frac{i}{8}\left(\frac{2\pi}{k}\right) [w{\bf D}_+\bar {\bf Z} +{\bar w}{\bf D}_+ {\bf Z} -    w\bar{\bf Z}   -  {\bar w} {\bf Z},{\bf A}_+] 
\ .
 \end{align}
The rest of the equations remain unchanged. In particular the Hamiltonian is the same except for an additional term in ${\cal P}_+$:
\begin{align}
\pi \int D_+Z_4  D_+ {\bar Z}_4  = \pi \int dzd\bar z \partial_+w\partial_+{\bar w}\ .
\end{align}
This will diverge unless $\partial_+w =  0$ as $w$ is holomorphic (although it would be finite for constant $w$ if we are on a compact Riemann surface).

Lastly we can  quantize the system in a natural way by considering wavefunctions $\psi(\xi^m)$ and replacing 
\begin{align}
\partial_+ \xi^m\to -i\frac{\partial \psi}{\partial \xi^m}\ .
\end{align}
Thus the dynamics reduces to quantum mechanics on Hitchin moduli space.

\section{Physical Interpretation}

So far in this paper we have solved the constraints of the $(2,0)$ superalgebra of \cite{Lambert:2016xbs} for a particular choice of three-form $C=l^3 dx^3\wedge dx^4\wedge dx^+$. We showed that the resulting system of equations had a vacuum configurations consisting of solutions to the Hitchin system on ${\bf R}^2$. We also saw that  the dynamical evolution  consisted of motion on the moduli space ${\cal H}_K(su(2),{\mathbb R}^2)$ of such solutions. Here  ${\cal H}_n({\mathfrak g},\Sigma)$ denotes the moduli space of the charge $n$ Hitchin system with gauge algebra ${\mathfrak g}$ on a Riemann surface $\Sigma$.  Therefore it is of interest to see how our construction fits in with other known descriptions of M-branes.

To begin with we recall that to solve the constraints of the original $(2,0)$ algebra we had to dimensionally reduce the full six-dimensional system on $x^3,x^4$ and $x^-$. However it is clear from the subsequent analysis that the resulting system still carries information about the momentum around $x^-$ in the form of the topological term ${\cal W}\sim \int T_{--}$. Thus we should view the system as two  M5-branes compactified on ${\mathbb T}^2\times S^1_-$ but with a fixed null momentum ${\cal P}_-\sim {\cal W}$.  

We can view  a null compactification as a limit of a boosted spacelike compactification where $x^5$ is taken to be compact with a radius that vanishes so that in the limit of a null boost the radius $R_-$ remains finite.
Therefore let us review the case where $C =l^3 dx^3\wedge dx^4\wedge dx^5$ is spacelike and the constraints imply that the fields have no dependence on $x^3,x^4,x^5$. It was shown in  \cite{Lambert:2016xbs} that the  $(2,0)$ superalgebra reduces to the description of two M2-branes with a transverse ${\mathbb R}^8$.  From a brane perspective we can think of this as a toroidal compactification on $x^3,x^4,x^5$, sending all the radii to zero,  accompanied by a U-duality transformation which decompactifies the dual torus. This can be thought of as an M-theory version of T-duality that takes $N$ M5-branes wrapped on ${\mathbb T}^3$ to $N$ M2-branes which are transverse to a dual $\hat {\mathbb T}^3$.\footnote{For the sake of generality here we have considered an arbitrary number of M-branes whereas the results we found above only concern the case of $N=2$.} In particular the U-duality  we require consists of reducing to string theory on $x^5$, leading to $N$ D4-branes wrapped on a $ {\mathbb T}^2$ with a coupling $g^2_{YM}\sim R_5$, and then performing T-dualities along $x^3$ and $x^4$ 
to find $N$ D2-branes with a transverse $\hat {\mathbb T}^2\times {\mathbb R}^5$ where the radii are $\hat R_3 =\alpha'/R_3$ and $\hat R_4 =\alpha'/R_4$ and  the coupling constant is  $\hat g^2_{YM} \sim R_5/R_3R_4$. If we now shrink the original radii to zero we obtain the strong coupling limit of $N$ $D2$-branes in a transverse ${\mathbb R}^7$ or equivalently  $N$ M2-branes in a transverse ${\mathbb R}^8$.

\hskip-1cm\includegraphics[scale=.43,angle=0]{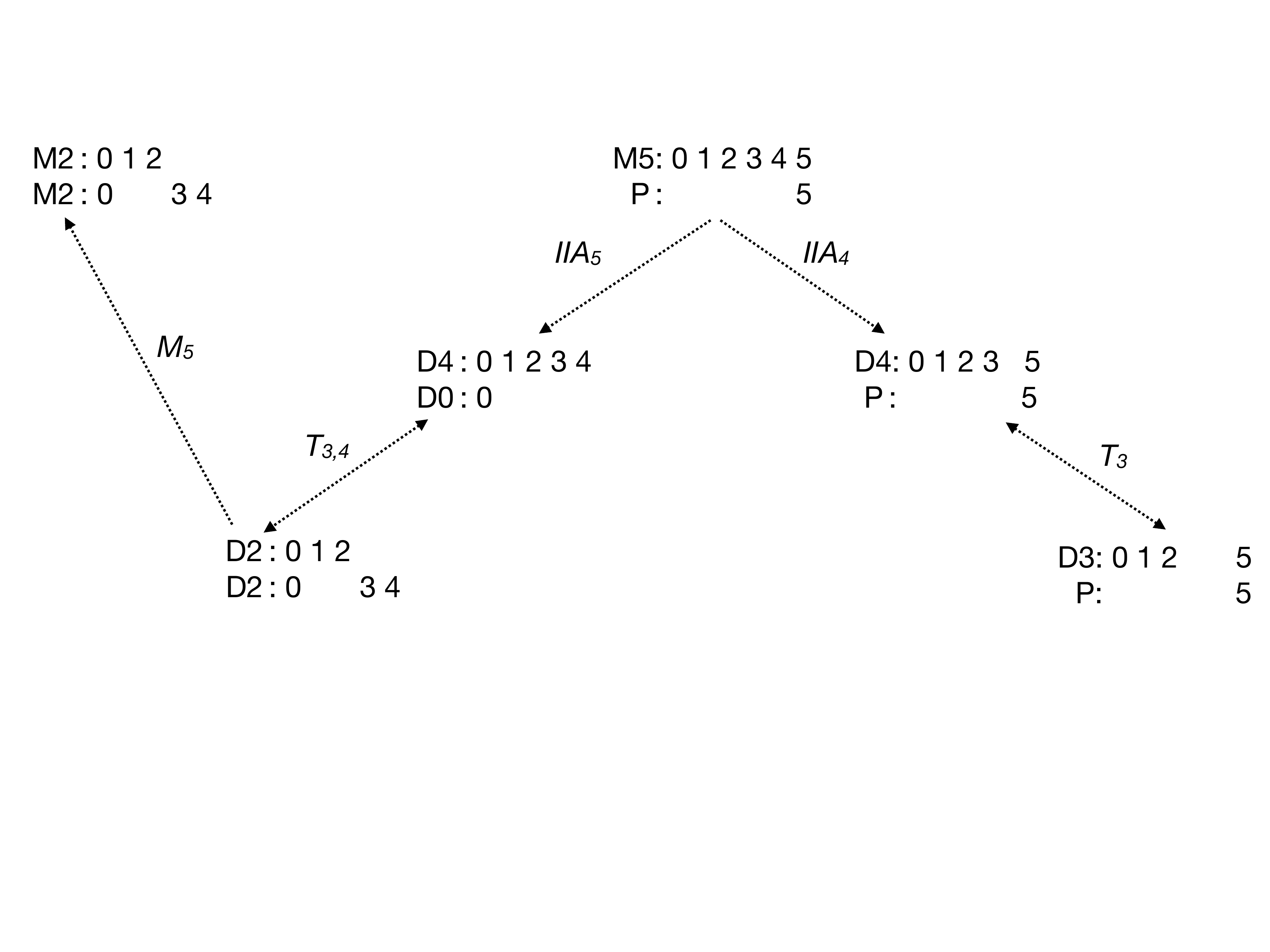} 
\vskip-4cm \noindent{{\bf Figure 1:} U-dualities of an M5 with momentum. $IIA_n$ indicates reduction to string theory along $x^n$, $T_n$ T-duality along $x^n$ and $M_n$ lift to M-theory along $x^n$.  }

Let us repeat these steps with $K$ units of momentum along $x^5$. In addition to the $N$ D4-branes we also find $K$ D0-branes. After T-duality these become $K$ D2-branes along $x^3,x^4$. Taking all the radii to zero leads to $N$ M2-branes along $x^0,x^1,x^2$ and $K$ M2-branes along $x^0,x^3,x^4$. The Hitchin system can then be thought of as the BPS condition for $K$ M2-branes intersecting the original $N$ M2-branes, generalising the familiar abelian holomorphic condition $\bar \partial Z=0$ for intersecting branes. We also see that there will be an $SO_L(2)\times SO_R(2)\times SO_R(6)$ symmetry from rotations in the $(x^1,x^2)$, $(x^3,x^4)$ and $(x^5,...,x^{10})$ planes respectively. 

Lastly we need to perform the light-like boost along $x^5$ which is transverse to all the M2-branes. In terms of static gauge this corresponds to replacing $X^5$ with $-vx^0+ X^5$ and taking the limit $v\to 1$. For  $v \ne 0$ this will break the $SO_R(6)$ symmetry of the total transverse space to $SO(5)$. However one can see that  the breaking only occurs through the time derivative kinetic terms. The spatial gradient terms will remain invariant under $SO_R(6)$. The interaction terms also remain  invariant since $X^5\to -vx^0+X^5$ is  a shift by the centre of mass degree of freedom which is non-interacting\footnote{This is clear for D2-branes where the centre of mass degree of freedom is given by the identity matrix and all interactions are through commutators. This degree of freedom can be somewhat subtle in interacting M2-brane models but ultimately one expects this statement to remain true.}. If we take the limit $v\to 1$ then 
the M2-brane tension vanishes, the kinetic terms diverge and we are forced to set them to zero. Thus the $SO_R(6)$ symmetry is restored. In addition  we can allow the moduli to evolve such that $\partial_0\xi^m \sim {\cal O}(\sqrt{1-v^2})$. In this case the $SO_R(6)$ symmetry remains unbroken  as these moduli are invariant under rotations of the total transverse space. In the limit that $v\to 1$ the Manton approximation of slow motion on the moduli space of solutions becomes exact  and the dynamics reduces exactly to motion on ${\cal H}_K(su(2),{\mathbb R}^2)$.  

This agrees with the results that we have found in the previous section. Stated somewhat differently boosting the intersecting M2-branes leads to `fast' modes corresponding to the over-all transverse scalars $X^I$ (what we called the background fields before) and `slow' modes corresponding to the moduli $\xi^m$. Time evolution of the `fast' modes breaks   $SO_R(6)$ to $SO(5)$  but time evolution of the `slow' modes does not. Thus the   $(2,0)$ system we obtained above can be viewed as describing the `slow' modes,  with the `fast' modes   frozen or integrated out ({\it i.e.} set to their expectation values).  


Let us now comment on a separate but related description of $N$ M5-branes on ${\mathbb T}^2\times S^1_-$. In particular let us first compactify on ${\mathbb T}^2$. As is well known reduction of the $A_{N-1}$ $(2,0)$ theory on a torus of vanishing area (but fixed shape) leads to maximally supersymmetric $U(N)$ Yang-Mills. More precisely we can reduce to string theory on $x^4$ to obtain $N$ D4-branes with coupling $g^2_{YM}\sim R_4$ and then T-dualise along $x^3$ to find $N$ D3-branes with finite coupling  $g^2 \sim R_4/R_3$.
Lastly we introduce $K$ units of null momentum along $x^5$ which leaves a manifest $SO(2)\times SO(6)$ symmetry that arises from rotations in the $(x^1,x^2)$ and $(x^3,x^6,x^7,x^8,x^9,x^{10})$ planes respectively. This is the set-up for a  DLCQ construction of four-dimensional maximally super-symmetric Yang-Mills.   This was given in \cite{Kapustin:1998pb} in terms of the quantum mechanics on ${\cal H}_N(u(K),\hat {\mathbb T}^2)$ where $\hat {\mathbb T}^2$ is an auxiliary two-torus.    Various details of this system have been studied in detail more recently in \cite{Dorey:2014dza} and see also \cite{Ganor:1997jx} for an alternative description.

These two descriptions differ by a T-duality along $x^4$ as well as a U-duality 
corresponding to the choice of M-theory direction (a `$9-11$ flip' that swaps $x^4$ with $x^5$). However it is also possible that the two descriptions involve different choices of `fast' and `slow' modes.  In the case of D3-branes there is a manifest $SO(2)\times SO(6)$ symmetry that comes from rotations in the $(x^1,x^2)$ and $(x^3,x^6,x^7,x^8,x^9,x^{10})$ planes respectively. In the case of M2-branes we saw that there is an $SO(2)\times SO(2)\times SO(6)$ symmetry corresponding to rotations in the $(x^1,x^2)$ and $(x^3,x^4)$ and $(x^5,x^6,x^7,x^8,x^9,x^{10})$  planes respectively. This enhancement of the R-symmetry from $SO(2)\times SO(6)$ to $SO(2)\times SO(2)\times SO(6)$ presumably comes from taking the strong coupling limit corresponding to the lift to M-theory. Therefore we expect it to be present in the strong coupling DLCQ description of D3-branes but only in the case where $R_3=R_4$.

Perhaps a more direct relation between the two descriptions can been seen as follows. We are free to compactify ${\mathbb R}^2$ to a torus ${\mathbb T}^2_{12}$.  Our M2-brane description then becomes motion on ${\cal H}_K(su(N),{\mathbb T}^2_{12})$ and the   $SO_L(2)\times SO_R(2)\times SO_R(6)$ symmetry is broken to  $SO_R(2)\times SO_R(6)$.  If we reduce to string theory on $x^5$ we again  obtain the intersecting D2-branes discussed above but we can now T-dualise along $x^1,x^2,x^3,x^4$ and then lift back to M2-branes. This has the effect of simply swapping the original $N$ M2-branes that were tangent to $x^0,x^1,x^2$ with the $K$ intersecting M2-branes that were tangent to $x^0,x^3,x^4$. The result is  motion on ${\cal H}_N(su(K),\hat {\mathbb T}^2_{12})$ where  $\hat {\mathbb T}^2_{12}$ is the T-dual torus to ${\mathbb T}^2_{12}$. This is almost in agreement with the DLCQ description if we identify $\hat {\mathbb T}^2$ with $\hat {\mathbb T}^2_{12}$. However there is one caveat: we see only the $su(K)$ Lie algebra and not $u(K)$. We assume that this came about because of the gauge group of the three-algebra associated with maximal supersymmetry is $su(2)\oplus su(2)$ rather than $u(N)\oplus u(N)$ that arises in the ABJM model. Thus it would seem that the T-duality and U-duality discussed  above manifest themselves as a rank-charge duality in the quantum mechanics on the Hitchin moduli space.

Lastly let us examine the formula for ${\cal W}$ in the case that we considered in section 3.3 and propose an interpretation for it as the M5-brane momentum ${\cal P}_-$.  It is known that there are no finite action regular solutions to the Hitchin system on ${\mathbb R}^2$ \cite{Saclioglu} (more recently see  \cite{Ward}) but here we will make a proposal on how to interpret certain multi-valued solutions.
Restoring the factor of $l$, identifying $\langle A,B\rangle = -\frac12{\rm tr}({\bf A}{\bf B})$ (valid in the case considered in section 3.3) and replacing the integral over $x^3,x^4,x^-$ by the volume factor 
$V_3=(2\pi)^3R_3R_4R_-$ that we would get by taking $x^3,x^4,x^-$ to be periodic we have
\begin{equation}
{\cal W} = \frac{\pi  }{2l^6}V_3\frac{i}{2}\int d{\rm tr}(\bar {\bf Z}{\bf D}{\bf Z}dz)  - d{\rm tr}({\bf Z}\bar {\bf D}\bar {\bf Z}d\bar z) \ .
\end{equation}
For a smooth solution the integral is only over the sphere at infinity. Let us assume
that for large $z$ we can treat ${\bf Z}$ as abelian and ignore ${\bf A}$ (which can either be subleading or simply commuting with ${\bf Z}$). Then up to a gauge transformation we can expand
\begin{equation}
{\bf Z} =  -i{a}    {\bf J}_3\ln z + {\bf C} + \dots\ ,
\end{equation}
where ${\bf J}_3$ is a real anti-hermitian generator of ${so}(3)$ normalised to ${\rm tr}({\bf J}_3^2)=-2$ and the ellipsis denotes subleading terms. We have assumed this asymptotic form so that ${\cal W}\ne 0$. Even so the expression for ${\cal W}$ is  problematic as there is a divergence:
\begin{align}
{\cal W} = &-\frac{\pi i}{4l^6}V_3\left[\oint 2|a|^2 \frac{\ln \bar z}{z}dz +i   {\rm tr}(a{\bf J}_3\bar {\bf C})\oint\frac{1}{z} dz \right]+c.c.\ .
\end{align}
 However if we cut-off the divergent terms at some large by finite $r=|z|$ they become \begin{align}
{\cal W}_\infty & = -\frac{\pi i}{4l^6}V_3a|^2\oint \frac{\ln \bar z}{z} dz +c.c. \nn\\
  & =- \frac{\pi i}{4l^6}V_3|a|^2\oint {\ln \bar z}  d\ln z +c.c.\nn\\
& = -\frac{\pi i}{4l^6}V_3|a|^2\int_{ \ln r -i\pi}^{ \ln r+i\pi} \bar w d w +c.c.\nn\\
& =  \frac{\pi   }{4l^6}V_3|a|^2\int_{-\pi}^{\pi} (\ln r -i\theta) d\theta +c.c. \nn\\
& = \frac{ \pi^2  }{ l^6}V_3 |a|^2\ln r  \ ,
\end{align}
where we have introduced a branch cut for $\ln z$ that runs along the negative real axis and written $w=\ln r + {i\theta}$. 
Therefore we find
 \begin{equation}
 {\cal W} = {\cal W}_\infty+  \frac{\pi^2 i}{ 2l^6}V_3  \tr({\bf J}_3 (a\bar{\bf C}-\bar a{\bf C}))\ .
 \end{equation}
 
Next we observe that ${\bf Z}$ is not single valued: under a rotation $z\to e^{2\pi i}z$ we see that  ${\bf Z}\cong {\bf Z} + 2\pi    a {\bf J}_3 $. We recall that ${\bf Z} = {\bf Y}^4+i{\bf Y}^3$ where ${\bf Y}^4$ and ${\bf Y}^3$ are real anti-symmetric matrices. These have imaginary eigenvalues $y_4$ and $y_3$ respectively which, after multiplication by $i$, can be thought of as positions of the two M5-branes along $x^4,x^3$ directions. The above identification then implies that $y^4\cong y^4 + 2\pi {\rm Re} a$ and $y^3\cong y^3 + 2\pi{\rm Im} a$. We learn from this that $Y^3$ and $Y^4$ must be treated as periodic and hence we identify $a =   R_4+ iR_3$.  

This means that  the divergent term  only depends on $R_3,R_4, R_-$.  Unfortunately we do not have a physical interpretation for this divergence, it would be interesting to find one. However in this discussion we only want to consider solutions that correspond to fixed radii and so we will simply ignore the divergence and consider instead
 \begin{equation}
 {\cal W}_{finite} =    \frac{\pi^2i}{ l^6}V_3  \tr({\bf J}_3 (a\bar{\bf C}-\bar a{\bf C}))\ .
 \end{equation}
Let us write ${\bf C}=c{\bf J}_3+...$ where the ellipsis denotes terms that are orthogonal to ${\bf J}_3$. Thus
\begin{equation}
{\cal W}_{finite} =  -\frac{2\pi^2i}{ l^6}V_3  (a\bar c-\bar ac)\ .
 \end{equation}
 The multivalued nature of ${\bf Z}$ also means that in the space of solutions, those which differ by $c\to c+2\pi  a$ must be identified with each other.  Therefore if we write
\begin{equation}
c =  2\pi   R_4n_4+ 2\pi iR_3n_3 \ ,
\end{equation}
then solutions that differ by $(n_3,n_4)\to (n_3+1,n_4+1)$ are identified with each other. As a result we have
\begin{align}
{\cal W}_{finite} &=  \frac{8\pi^3}{l^6}V_3 R_3R_4(n_4-n_3) \nn\\
&=  \left(\frac{V_3}{l^3}\right)^2
\frac{n_4-n_3}{   R_-}\ .
\end{align}
This suggests that we should identify $l^3 =  V_3=(2\pi)^3R_3R_4R_-$ and so  recover the KK spectrum of a null compactification on $x^-$, provided that $n_4-n_3$ is an    integer. Putting this another way: in order to arrive at the interpretation of our model as describing a null compactification M5-branes we should assume $(Y^3,Y^4)$ are periodic and  impose on our Hitchin system the boundary condition ${\bf Z} \sim   -i(R_4+iR_3)    {\bf J}_3 ln z + 2\pi   (R_4n_4+   iR_3n_3 ) {\bf J}_3 $ where $n_4-n_3$ is an   integer. Lastly we mention that, according to the previous discussion, we are ultimately required to let $R_3,R_4,R_5\to 0$. However  when viewed as the limit of a null boost, the spacelike radius is sent to zero in such that a way that  $R_-$ is fixed. In this case ${\cal W}_{finite}$ remains finite. 
 
\section{Conclusion}

In this paper we presented a solution to the constraints of the $(2,0)$ system derived in \cite{Lambert:2016xbs}. The result was a system of equations for 3-algebra valued fields $Z,H,X^I,\Psi_\pm$, along with an associated gauge field one-form $A$, that are defined on a plane ${\mathbb R}^2$ times a null direction ${\mathbb R}_+$ which we used as `time'. We saw that for  choices of the fields $X^I,H,\Psi_-$ that preserve the ${\cal Q}_+$ supersymmetries the system reduced to  supersymmetric dynamics (with supersymmetry generator ${\cal Q}_+$) on the moduli space of an $SO(3)$ Hitchin system.
We also gave a physical interpretation of the resulting system as a re-formulation of the M5-brane on ${\mathbb T^2}\times S^1_-$ as  intersecting null M2-branes or alternatively a DLCQ of four-dimensional maximally supersymmetric Yang-Mills.

The original Hitchin system arises in our system for one particular choice of background. In addition our equations   admit generalizations such as a non-zero $Z_4$ and non-constant $X^I$. It would be interesting to examine these backgrounds and their associated dynamics. It is also possible to include impurities giving by sources in the Hitchin equations as done in \cite{Ganor:1997jx,Kapustin:1998pb}. We also expect that our results can be naturally extended to a Lorentzian 3-algebra and hence to an arbitrary gauge group. We also note that  Hitchin's system has also appeared before in conjunction with class-S theories derived from the M5-brane \cite{Gaiotto:2009hg,Yonekura:2013mya,Wang:2015mra,Xie:2014pua,Neitzke:2014cja}.

Lastly we note that the Hitchin system is generally thought of as applying to  a Riemann surface $\Sigma$ of genus $g$. However here we have taken the coordinates $(z,\bar z)$ to be those of the  flat plane, or possibly a torus, which admit covariantly constant spinors. Due to the $SO_R(2)$ symmetry we may twist our theory   by taking Killing spinors of the diagonal group of $SO_L(2)\times SO_R(2)$.  Alternatively we could break the transverse $SO(5)\to SO(3)\times SO(2)$ and use the later to twist the theory. Thus we expect to be able to extend our supersymmetric system to a generic Riemann surface and possibly make contact with the class-S theory literature (or at least toroidal compactifications of them). In doing so we should also allow for singularities   at marked points on the Riemann surface.     

\section*{Acknowledgements}
P. Kucharski would like to thank King's College London for its hospitality and is supported by ERC Starting Grant no. 335739 ``Quantum fields and knot homologies" funded by the European Research Council under the European Union's Seventh Framework Programme.   N. Lambert was supported in part by STFC grant ST/L000326/1 and M. Owen is supported by the STFC studentship ST/N504361/1.  

\section{Appendix: Conventions}

In the text we introduced the coordinates
\begin{align}
x^{+}&=\frac{x^{5}+ x^{0}}{\sqrt{2}}\qquad x^{-}=\frac{x^{5}- x^{0}}{\sqrt{2}}\ .
\end{align}
In these coordinates we find
\begin{align}
\eta_{+-}&=\eta_{-+}=1\nonumber \\
\epsilon_{1234+-}&=\epsilon_{+-1234}=-1\ .
\end{align}
For spinors we  find it useful to introduce the following conventions:
\begin{align}
\Gamma_{\pm} & = \frac{\Gamma_5 \pm \Gamma_0}{\sqrt{2}}\nn\\
\Gamma_{05} & = \Gamma_{+-}\ .
\end{align}
We then find that
\begin{align}
\Gamma_- \chi = \Gamma _- \chi_+ & = -\sqrt{2} \Gamma_0 \chi_+ \nonumber \\
\Gamma_+ \chi = \Gamma _+ \chi_- & = \sqrt{2} \Gamma_0 \chi_- \nonumber \\
\Gamma_{\pm}\chi_{\pm} & = 0\nn\\
\Gamma_-\Gamma_+\chi & = 2\chi_-\nn\\
\Gamma_+\Gamma_-\chi & = 2\chi_+\ .
\end{align}

We also introduced complex coordinates
\begin{equation}
z=x^{1}+ix^{2}\ ,
\end{equation}
so that 
\begin{align}
g_{z \bar{z}} &= \frac{1}{2} \qquad \varepsilon_{-+z\bar{z}34} = \frac{i}{2} \nn\\
 D \equiv D_z &= \frac{1}{2}\left( D_1 - i D_2 \right) \qquad \bar{D} \equiv D_{\bar{z}} = \frac{1}{2}\left( D_1 + i D_2 \right)   \ . 
\end{align}
We also define
\begin{eqnarray}
\hat\Gamma_{z} = \frac{1}{2} (\hat\Gamma_1 - i\hat\Gamma_2) =\frac{1}{2} ( \Gamma_{01} - i \Gamma_{02})  \nonumber \\
\hat\Gamma_{\bar{z}} = \frac{1}{2}(\hat\Gamma_1 + i\hat \Gamma_2)= \frac{1}{2} ( \Gamma_{01} + i \Gamma_{02})\ .
\end{eqnarray}

Next we introduced the complex scalar
\begin{equation}
Z=Y^{4}+iY^{3}\ ,
\end{equation}
and 
\begin{eqnarray}
\hat\Gamma_{Z} = \frac{1}{2} (\hat\Gamma_3 - i\hat\Gamma_4)=\frac{1}{2} (\Gamma_{054} - i\Gamma_{053}) \nonumber \\
\hat\Gamma_{\bar{Z}} =\frac{1}{2}(\hat\Gamma_3 + i \hat\Gamma_4)= \frac{1}{2}(\Gamma_{054} + i \Gamma_{053})\ .
\end{eqnarray}

\bibliographystyle{utphys}
\bibliography{1706.00232v2REVISED}

\end{document}